\documentclass[%
 reprint,
 amsmath,amssymb,
prx,longbibliography
]{revtex4-2}
\usepackage[T1]{fontenc}
\usepackage{hyperref}
\usepackage{xcolor}
\usepackage{graphicx}% Include figure files
\usepackage{dcolumn}% Align table columns on decimal point
\usepackage{bm}% bold math
\usepackage{physics}
\begin{document}

\preprint{APS/123-QED}

\title{The NV centre coupled to an ultra-small mode volume cavity: \\ A high-efficiency source of indistinguishable photons at 200 K}

\author{Joe A. Smith}\email{j.smith@bristol.ac.uk}
\affiliation{QET Labs, Department of Electrical and Electronic Engineering and H. H. Wills Physics Laboratory, University of Bristol, Bristol BS8 1UB, UK}

\author{Chloe Clear}
\affiliation{QET Labs, Department of Electrical and Electronic Engineering and H. H. Wills Physics Laboratory, University of Bristol, Bristol BS8 1UB, UK}

\author{Krishna C. Balram}
\affiliation{QET Labs, Department of Electrical and Electronic Engineering and H. H. Wills Physics Laboratory, University of Bristol, Bristol BS8 1UB, UK}

\author{Dara P. S. McCutcheon}
\affiliation{QET Labs, Department of Electrical and Electronic Engineering and H. H. Wills Physics Laboratory, University of Bristol, Bristol BS8 1UB, UK}

\author{John G. Rarity}
\affiliation{QET Labs, Department of Electrical and Electronic Engineering and H. H. Wills Physics Laboratory, University of Bristol, Bristol BS8 1UB, UK}

\vspace{5mm} %5mm vertical space


\begin{abstract}

Solid state atom-like systems have great promise for linear optic quantum computing and quantum communication but are burdened by phonon sidebands and broadening due to surface charges. Nevertheless, coupling to a small mode volume cavity would allow high rates of extraction from even highly dephased emitters. We consider the nitrogen vacancy centre in diamond, a system understood to have a poor quantum optics interface with highly distinguishable photons, and design a silicon nitride cavity that allows 99\% efficient extraction of photons at 200 K with an indistinguishability of $>$ 50\%, improvable by external filtering. We analyse our design using FDTD simulations, and treat optical emission using a cavity QED master equation valid at and beyond strong coupling and which includes both ZPL broadening and sideband emission. The simulated design is compact ($<$ 10 $\mu$m), and owing to its planar geometry, can be fabricated using standard silicon processes. Our work therefore points towards scalable fabrication of non-cryogenic atom-like efficient sources of indistinguishable photons.

\end{abstract}

\maketitle

%\tableofcontents

% PACS
% 42.50.Ct	Quantum description of interaction of light and matter; related experiments
% 2.50.Ex Optical implementations of quantum information processing and transfer

%-------------------------------------- Background ----
\section{\label{sec:level1}Introduction}
The nitrogen vacancy (NV) centre in diamond is perhaps the best understood optically-addressable atom-like system with a long-lived electron spin at room temperature, and is a promising platform for implementing quantum networks \cite{rozpkedek2019near}. Unfortunately, at room temperature ($300\,$K), the $15$ THz broadened zero-phonon line (ZPL), and a sideband which redshifts $>$ 95 \% of emission, means the native defect has a poor quantum optics interface with a very low probability of emitting single photons into single spatial and spectro-temporal modes. Such interfaces are essential for building future quantum networks as these `indistinguishable' photons can be used to entangle remote NV centre spins through high fidelity quantum interference \cite{bernien2013heralded}. The indistinguishability $I$ can be estimated by considering the natural linewidth $\gamma = 30$ MHz over the broadened linewidth $\gamma^*$ = 15 THz, scaled by a 0.02 Debye-Waller factor $D_W$ \cite{albrecht2013coupling}, which gives \cite{iles2017phonon}: 
\begin{align}\label{eq:bare}
    I = D_W^2\frac{\gamma}{\gamma + \gamma^*},
\end{align}
yielding $I = 0.8\times 10^{-9}$, effectively useless for generating remote entanglement events. Cooled to 4 K, the ZPL emission narrows dramatically with $\gamma^*$ reaching of order 1 GHz in nanostructured diamond \cite{riedel2017deterministic}, with Fourier-limited linewidths reported in bulk diamond \cite{chen2017laser}. Typically, the emitter is coupled to a high $Q$ cavity to enhance emission into the ZPL mode via the Purcell factor, with experiments currently reporting a ZPL fraction of 0.46~\cite{riedel2017deterministic}.  From this, spectral filtering the ZPL results in a quantum resource in the form of on-demand solid state sources of indistinguishable photons from which two or more can be interfered for generating entanglement over distances~\cite{bernien2013heralded,kalb2017entanglement}. 

Liquid helium cooling, although convenient in a laboratory experiment, brings complexity and constraints to single photon sources that are prohibitive for large-scale applications, especially as the number of sources exceeds ten. As such, improving operation to the regime of Peltier cooling ($200\,$K) or even room temperature would be highly beneficial~\cite{bogdanov2019overcoming,bogdanov2020ultrafast}. Additionally, it has been proposed that Purcell-enhancement could lead to single shot electron-spin readout by enhancing the room temperature spin contrast ~\cite{wolf2015purcell,jung2019spin}. Marrying these two ideas could eventually lead to spin-photon interfaces at elevated temperatures.

Further, the engineering of cavities that can overcome highly thermally dephased emitters has implications for other dephased solid state systems. Although cryogenic emission is spectrally narrow, wander due to charge noise remains problematic, resulting in a broader effective linewidth $\gamma^*$, especially as the NV centre is brought close to the surface to improve its photonic interface \cite{riedel2017deterministic} and the requirement for pre-selecting lifetime-limited NV centres severely restricts scalability.

Recently ultra-small mode volume photonic crystal cavities (PhCCs) in silicon waveguides have been designed and fabricated \cite{hu2018experimental}. These `bow tie' cavities would show extremely high spontaneous emission enhancements at reasonably low quality factors suitable for enhancing broader ZPL modes. Here, we demonstrate such a cavity design in silicon nitride, compatible with low fluorescence silicon nitride films encapsulating nanodiamond containing single NV centres \cite{smith2019single}. With an ultra-small ($ << \lambda^3$) mode volume, it is possible to achieve a large atom-cavity coupling rate $g$, which serves to increase the Purcell-enhanced spontaneous emission rate $R$ $\sim g^2/\kappa_c$, where $\kappa_c$ is the cavity linewidth (or leakage rate) constrained to be $\kappa_c < \gamma^*$. Increasing $R$ to the rate $\gamma^*$ results in the potential for indistinguishable photons to be extracted from the NV centre above cryogenic temperatures, despite its broad ZPL linewidth, as the emission leaves the atom-like environment faster than dephasing processes.

We model the ZPL of the NV centre coupled using a Lindblad master equation~\cite{kaer2013role,wein2018feasibility} to capture cavity regimes at the onset of strong coupling, which we then extend to consider the implications of the broad NV centre sideband on source figures of merit~\cite{iles2017phonon}. The sideband has a significant effect on the indistinguishability of photons from the NV centre, and we observe a trade-off between cavity-filtration of sideband photons and avoiding strong coupling, which causes the emission to split due to vacuum Rabi oscillations. We show that high indistinguishability can be observed at $200\,$K, especially when externally filtered to remove the remaining sideband fraction. Finally, we use the parameters from this study to simulate an ultra-small mode volume cavity, exploring how fabrication tolerances would affect the cavity performance.

\begin{figure}
\includegraphics[width=\columnwidth]{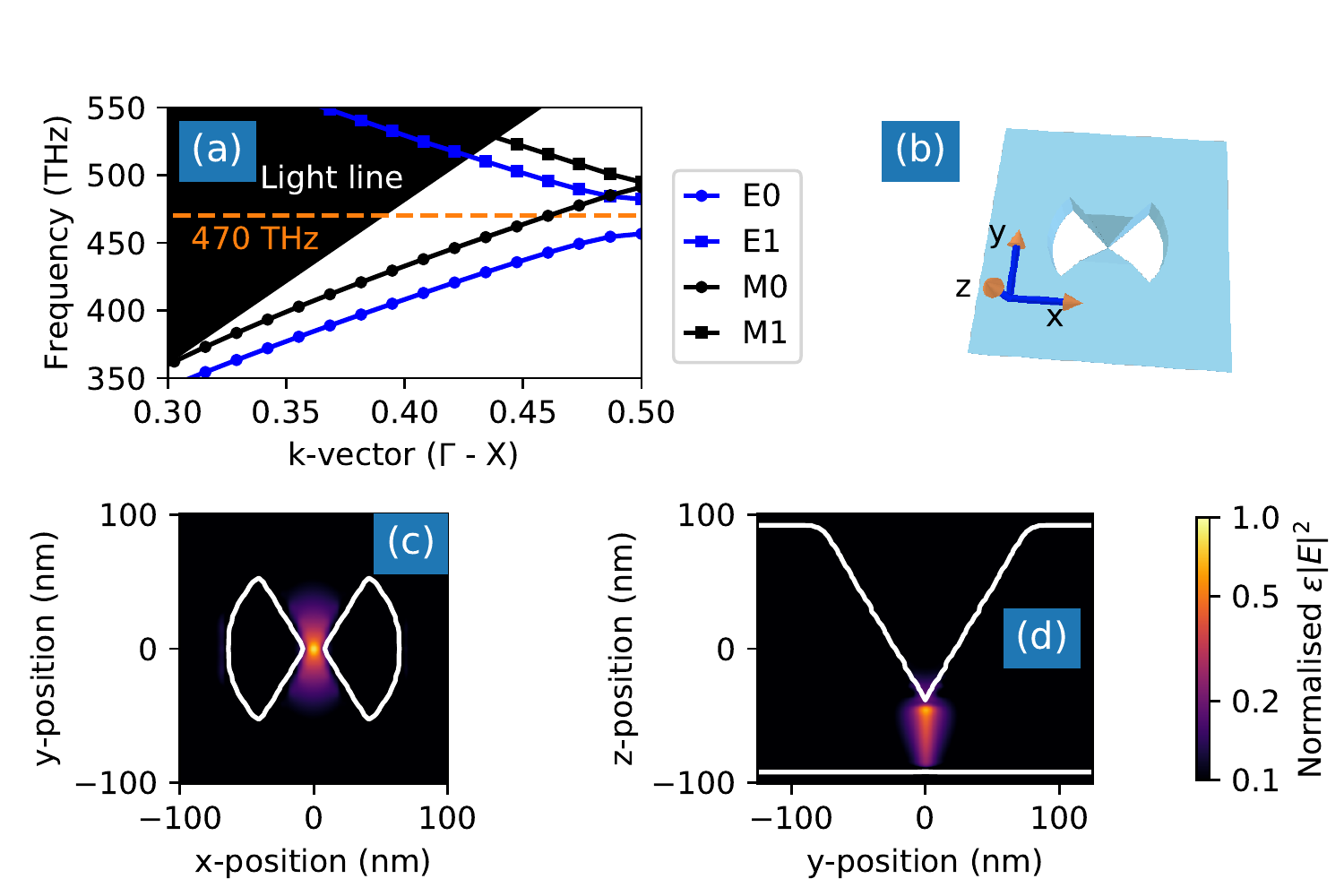}
\caption{\label{fig:phc} (a) Bandstructure of a silicon nitride bow tie photonic crystal \cite{hu2018experimental} designed at $\lambda = 637$ nm around the TE mode. (b) Photonic crystal unit cell showing bow tie confinement and V-groove z confinement. Electrical energy density $|E|^2$ at the band edge shows the ultra small mode volume realised (c) by the bow tie in the plane and (d) a non-planar incision into the dielectric at x = 0. }
\end{figure}

\section{Engineering large coupling from small mode volumes}
In cavity QED, the coupling rate $g$, which characterises the interaction of a dipole with the electric field operator, scales with the field mode volume $V_m$ as $g \sim 1/\sqrt(V_m)$. In weak coupling, the rate of transfer from the atom to external photons $R$ scales as $g^2/\kappa_c$. Where dephasing is significant, as in the NV centre, the interaction strength $g$ must be strong to overcome the dephasing rate $\gamma^*$ to produce useful indistinguishable photons, and we require $g^2/\kappa_c \geq \gamma^*$. Wavelength-scale mode confinement, with $V_m \sim (\lambda/n)^3$, was a key advance in PhCCs a decade ago. Sipahigil \emph{et al.} demonstrated that, with $V_m = 2.5(\lambda/n)^3$, they obtained a coupling of $g = 13$ GHz to a silicon vacancy in diamond \cite{sipahigil2016integrated}. Recently, Hu \emph{et al.} fabricated a PhCC with a mode volume of $ V_m \sim $ 0.001 $(\lambda/n)^3$~, measured as the integral of the electrical energy density normalised by the maximum energy density \cite{hu2018experimental}:
\begin{align}
     V_m =\frac{1}{\max(\varepsilon E^2)} \int \varepsilon E^2dV,
\end{align}

This small mode volume was achieved by modifying a standard cylinder photonic crystal cell to contain a `bow tie' of high index material leading to the mode being concentrated at the apex of the bow tie. Further confinement in the out of plane z-direction was achieved by thinning the centre of the bow tie by etching a V-groove. We model a bow tie photonic crystal unit cell in a 200 nm thick silicon nitride ($n = 2.0$) waveguide using Lumerical FDTD. In Fig.~\ref{fig:phc}(a) we demonstrate that a photonic crystal bandgap forms around the ZPL linewidth at 637 nm (470 THz), by repeating 250 nm unit cells, from which a defect can be introduced to realise a cavity \cite{hu2018experimental}. This single mode has a TE polarisation with the dielectric band (labelled E0) at 450 THz and the dielectric band (labelled E1) at 490 THz and would filter a large region of the broad sideband compared to the narrow free spectral range of open-access or ring-like microcavities. We also find the first TM-like band at 500 THz (between M0 and M1), above the NV centre sideband energy. In Fig~\ref{fig:phc}(b), we illustrate the photonic crystal unit cell. Fig.~\ref{fig:phc}(c) illustrates the tightly confined electrical energy density $\varepsilon|E|^2$ of this structure in the plane, and Fig.~\ref{fig:phc}(d) as a cross-section, demonstrating sub-wavelength confinement is achievable in lower index contrast media. 

From the simulated field, this geometry has a mode volume $V_m$  = 0.0052 $(\lambda/n)^3$ at the ZPL of the NV centre which should provide significantly enhanced NV centre coupling. In the next section, we consider optimal cavity parameters ($g$,$\kappa$) rigorously, focusing on maximising indistinguishability and efficiency, and incorporate the effects of the broad NV centre phonon sideband in its single photon emission.

\section{\label{sec:i-b}Indistinguishability of spectrally broad single photons}
Single photons are useful for simple quantum information protocols such as encoding bases in quantum key distribution. However, for more challenging quantum information tasks it is important that 1) two photons from a single emitter are indistinguishable from each other, and 2) photon emission occurs with high efficiency, i.e.  generated on demand. To calculate the ZPL indistinguishability, we model the NV centre--cavity system as a two-level system with pure dephasing. This is coupled to a one sided single mode cavity with annihilation (creation) operator $a$ $(a^{\dagger})$, with the Lindblad master equation
\begin{align}
\partial_t \rho(t) =&-ig\qty[X,\rho(t)]\nonumber\\
&+ \gamma\mathcal{L}_{\sigma}[\rho(t)]+\gamma^* \mathcal{L}_{\sigma^{\dagger}\sigma}[\rho(t)]+\kappa_c \mathcal{L}_{a}[\rho(t)],
\end{align}
where $\kappa_c$ is the cavity width and $X=\sigma^{\dagger}a+\sigma a^{\dagger}$.
From this, we numerically integrate the two-colour emission spectra to find the indistinguishability of the ZPL 
\begin{align}
    I_{\mathrm{\mathrm{ZPL}}} = \frac{\int^\infty_{-\infty} d\omega \int^\infty_{-\infty}d\nu|\mathcal{S}_{\mathrm{\mathrm{ZPL}}}(\omega,\nu)|^2}{ \mathcal{F}_{\mathrm{\mathrm{ZPL}}}^2},
\end{align}
where $\mathcal{F}_{\mathrm{\mathrm{ZPL}}}=\int^\infty_{-\infty} d\omega  \mathcal{S}_{\mathrm{\mathrm{ZPL}}}(\omega,\omega)$  is the power into the ZPL. 

\begin{figure}[t]
\includegraphics[width=\columnwidth]{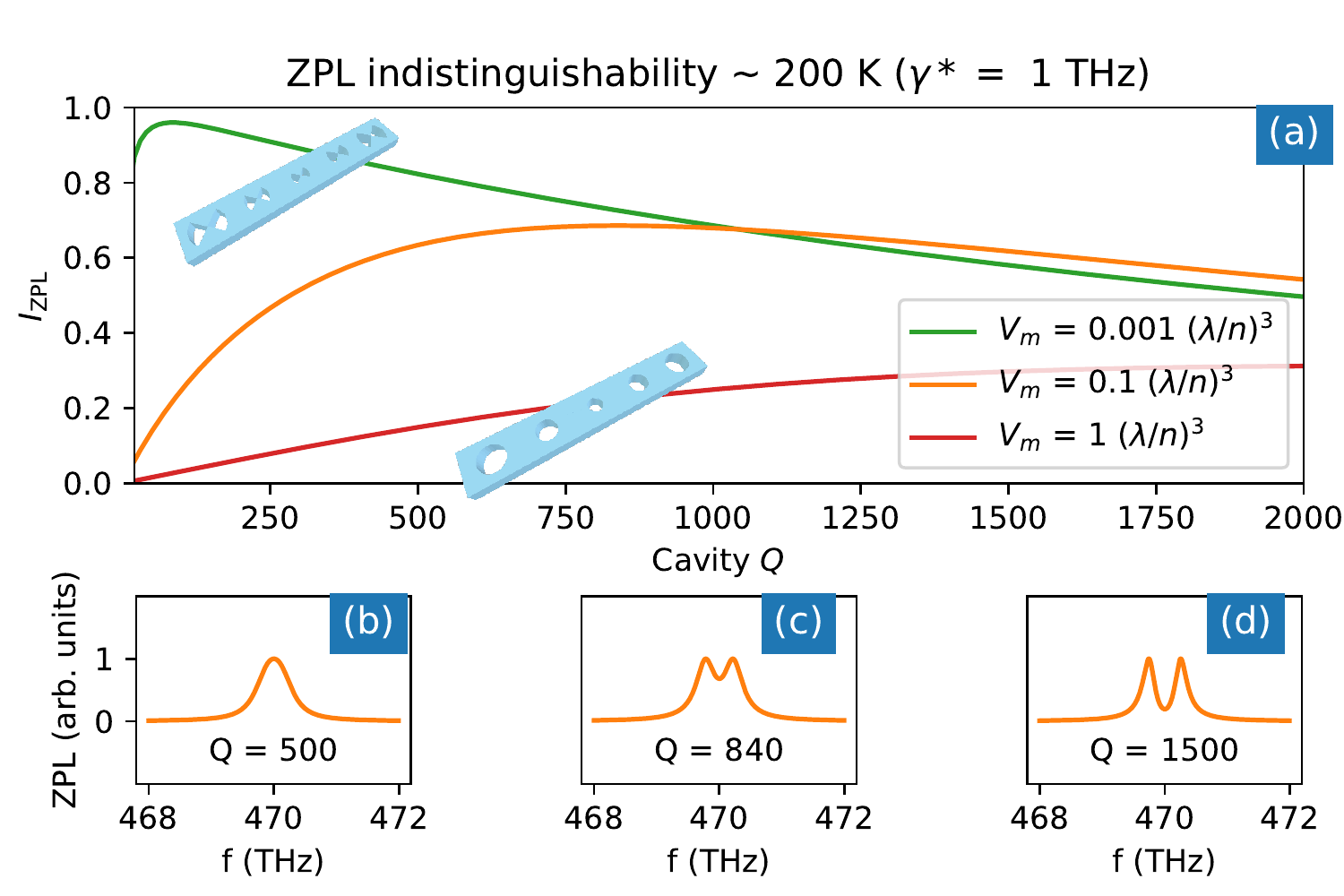}
\caption{\label{fig:inds} (a) Indistinguishability of the zero-phonon line ($I_\mathrm{ZPL}$) at 200 K. In red, cavities with a mode volume $V_m$ = 1 $(\lambda/n)^3$ cannot achieve high $I_\mathrm{ZPL}$. In green, the bow tie cavity with 0.001 $(\lambda/n)^3$ allows $I_\text{ZPL}$ = 0.96 although this requires a low $Q$ to prevent strong Rabi-induced splitting of the emission. (b)-(d) Effect of varying $Q$ on the 0.1 $(\lambda/n)^3$ cavity ZPL. Maximal $I_\mathrm{ZPL}$ occurs centre at slight splitting.}
\end{figure}

In Fig.~\ref{fig:inds}(a), we calculate $I_{\mathrm{\mathrm{ZPL}}}$ at different $V_m$ (see supplementary material for full details \footnote{See Supplemental Material for theoretical and simulation detail to support the main text.}) at large dephasing, $\gamma^*$ = 1 THz, present in non-cryogenic systems.  Parameterised by $(V_m, Q)$, the region of small $V_m$, low $Q$, affords high indistinguishability $I$ and cavity efficiency $\mathcal{F}_{\mathrm{\mathrm{ZPL}}}$ \cite{grange2015cavity}.  We observe that, in contrast to a bow tie design, the 1 $(\lambda/n)^3$ cavity cannot produce $I_{\mathrm{\mathrm{ZPL}}} > 0.4$. Arbitrary strong $g$ without $\kappa$ dampening causes the ZPL to be split, reducing indistinguishability, causing a point of optimal $I_{\mathrm{\mathrm{ZPL}}}$ at the onset of strong coupling where the emission splits into two modes (see Fig.~\ref{fig:inds}(b)-(d) ).

To capture the sideband emission of the NV centre we use a phenomenological approach with the sideband modelled as seven Lorentzian lines~\cite{albrecht2013coupling}. It is assumed that the sideband is incoherent and therefore gives rise to only distinguishable photon emission. To determine the impact of the sideband on our indistinguishability efficiency figures of merit, we assume that the cavity acts as a filter with width $\kappa_c$ on the off-resonant sideband emission, 
resulting in a sideband relative power $\mathcal{F}_{\mathrm{SB}}=\int^\infty_{-\infty} d\omega |h_c(\omega)|^2\mathcal{S}^{(0)}_{\mathrm{SB}}(\omega)$, where $h_c(\omega)=i\kappa_c/2(i\omega-\kappa_c/2)^{-1}$~\cite{iles2017phonon}. This gives an overall indistinguishability 
\begin{align}
I= I_{\mathrm{\mathrm{ZPL}}} \bigg(\frac{\mathcal{F}_{\mathrm{\mathrm{ZPL}}}}{\mathcal{F}_{\mathrm{\mathrm{ZPL}}}+\mathcal{F}_{\mathrm{SB}}}\bigg)^2,
\end{align}
with the two factors giving contributions arising from the ZPL and sideband respectively. Analytical expressions \cite{wein2018feasibility} for $I_{ZPL}$ were first used to find the optimal parameters with the final values of $I$ given from the numerical model. Further details of this model are given in the supplementary material \cite{Note1}.

Fig.~\ref{fig:sideband}(a) shows the calculated spectrum at 300 K, while in Fig.~\ref{fig:sideband}(b) we show the calculated photon indistinguishability, from which we see it is possible to obtain $I = 0.12$ with an appropriately designed bow tie cavity with parameters $V_m = 0.001(\lambda/n)^3$ and $Q = 200$. However, this broad cavity retains much of the incoherent sideband photons, and there is a trade-off for optimal $I$ between the Purcell enhanced spontaneous emission and the spectral filtering of the sideband. In comparison, in Fig.~\ref{fig:sideband}(c), at the limit of non-cryogenic cooling $T=200\,$K, the thermal broadened ZPL narrows to $\gamma^* = 1$ THz whilst the sideband remains broad \cite{fu2009observation}. The reduction in the $\gamma^*$ rate at 200 K allows $\kappa$ to be reduced without detrimentally affecting the indistinguishability. The resultant higher $Q$ leads to a significant percentage of the sideband to be filtered, as demonstrated in Fig \ref{fig:sideband}(d), reaching $I = 0.54$.

%In Fig \ref{fig:sideband}(d), with a narrower ZPL, we can operate in a higher $Q$ regime compared to 300 K \textcolor{red}{insert justification on trade-off of $\kappa$ and $\gamma^*$ to satisfy reviewer } and filter a significant percentage of the sideband reaching .

In both cases, $I$ can be further increased at the detriment of the system efficiency by externally filtering the emission with a Fabry--Perot filter with width $k_f$, as shown in the insets. Here the efficiency $\beta$ is the power into the filtered spectra as a proportion of the unfiltered spectra. In the 200 K system, this results in $I = 0.73$ at $\beta = 0.29$ with $\kappa_f = 0.3$ THz. Due to the assumption of a completely incoherent sideband, for very tight external filters ($\kappa_f <R\gamma$) our model likely underestimates the true value of $I$, as in this regime sideband photons themselves are expected to regain coherence~\cite{deng2019quantum}

\begin{figure}[b]
\includegraphics[width=\columnwidth]{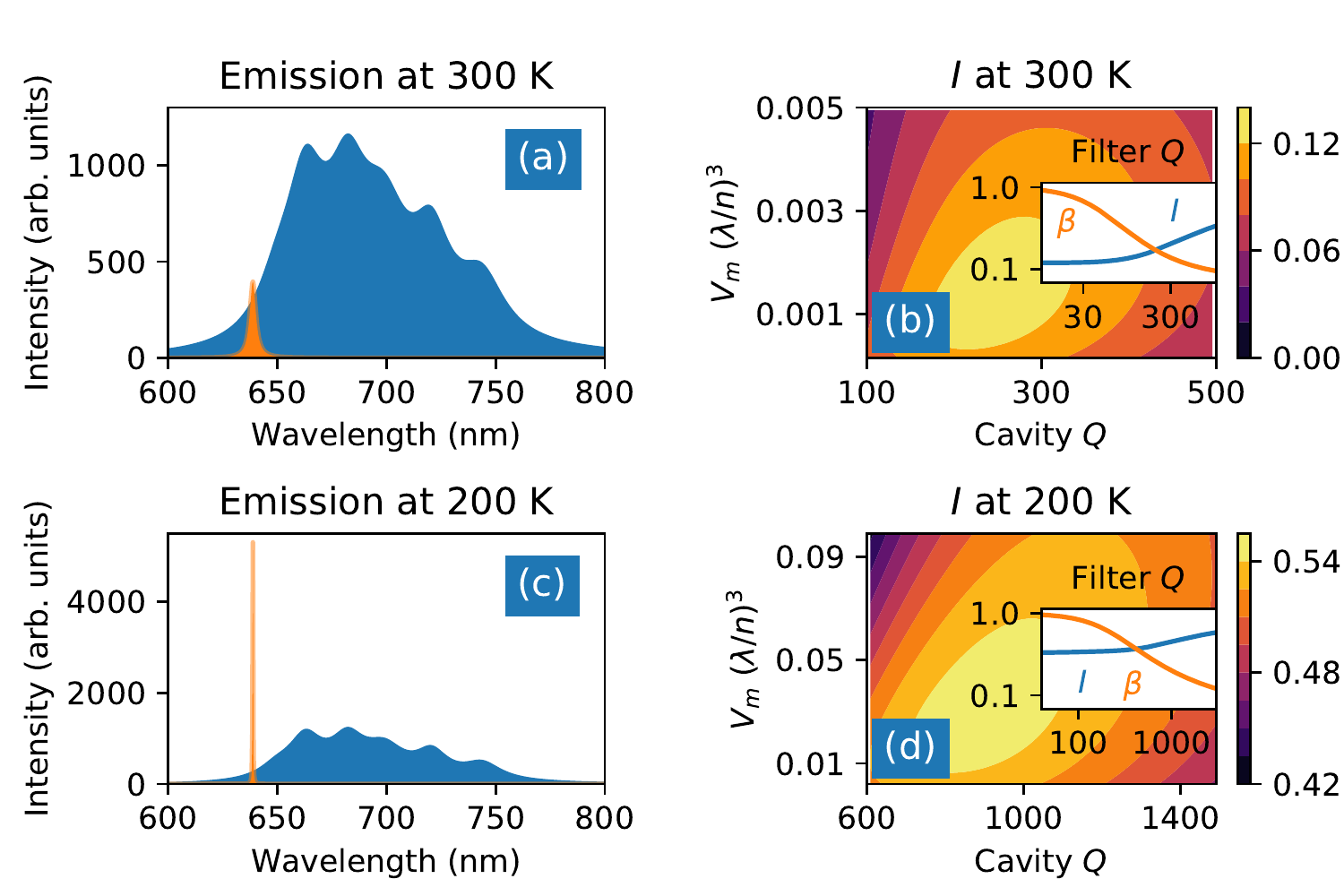}
\caption{\label{fig:sideband} Cooling the NV centre dramatically narrows the ZPL (in orange (a) and (c)). (b) To capture the ZPL at 300 K, low $I$ results because incoherent sideband photons are not filtered. (d) At 200 K, optimal $I = 0.54$ with more sideband filtered. Inset: an external filter can boost $I$ by reducing efficiency $\beta$ at the maximum of the contour.}
\end{figure}

\begin{figure}[b]
\includegraphics[width=\columnwidth]{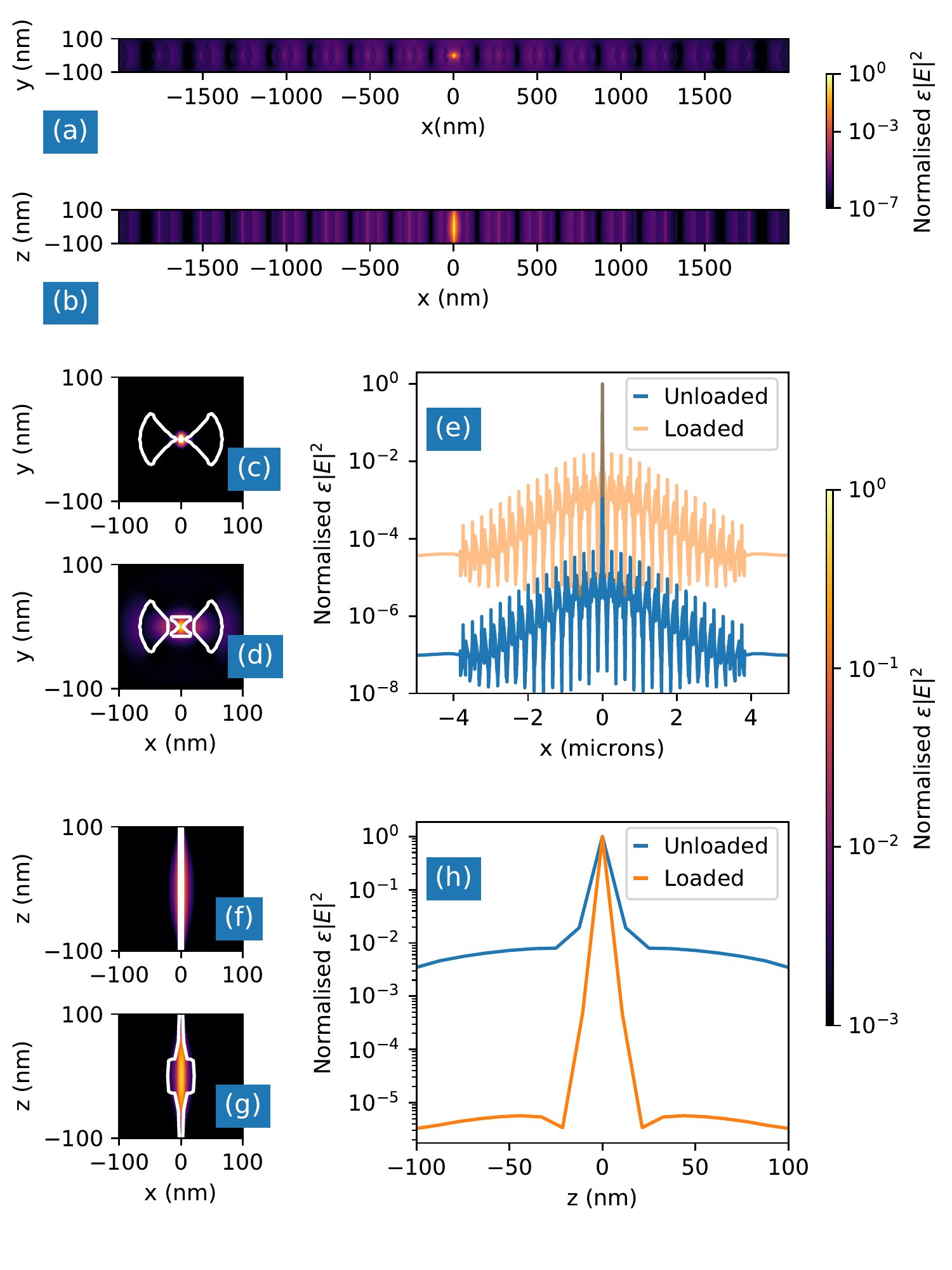}
\caption{\label{fig:cav} (a) Gaussian mode demonstrated by modulating the diameter of the circular bow tie along the waveguide. (b) The planar structure increases $V_m$ compared to Fig \ref{fig:phc}. When (c) the cavity centre is loaded with (d) a 20 nm cube to model the nanodiamond, (e) the x-plane modal confinement is diminished. (f) In contrast, the index contrast of (g) the nanodiamond improves (h) z-plane confinement.}
\end{figure}

\begin{figure}
\includegraphics[width=\columnwidth]{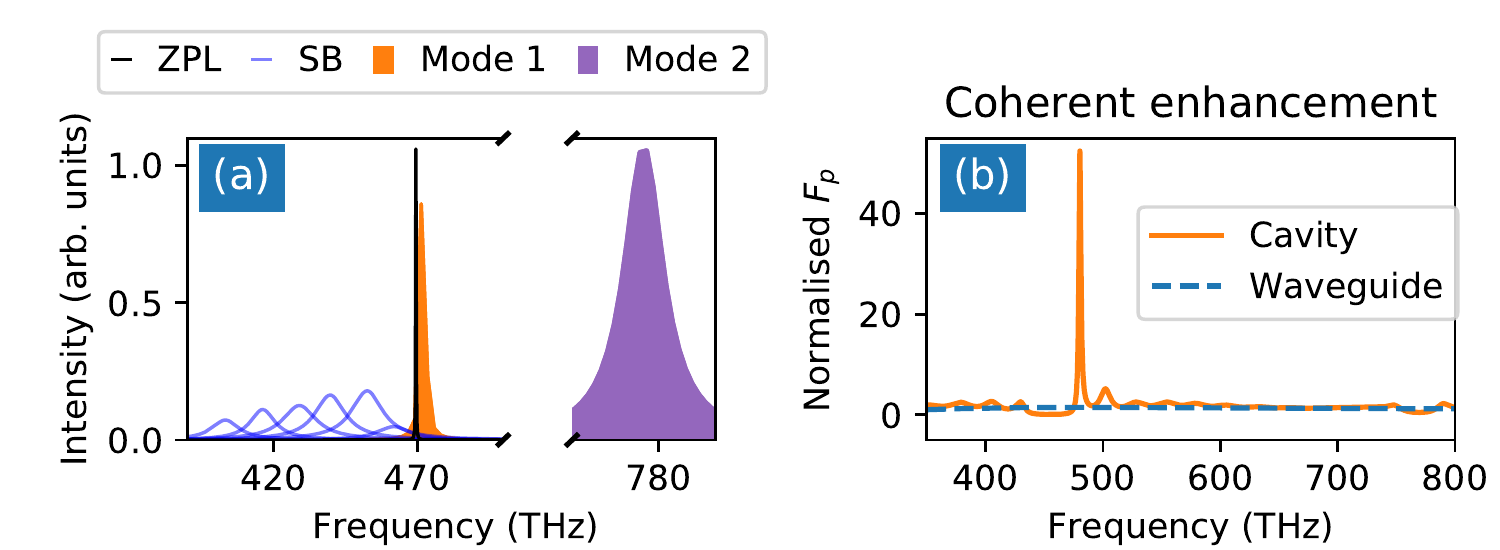}
\caption{\label{fig:cavspec}(a) Harmonic inversion of the simulated cavity response used to extract the resonances and $Q$, compared to the NV centre model. (b) Spectral response to a coherent dipole for the nanodiamond loaded cavity (orange) showing high enhancement at the first order mode and suppression in the band gap, compared to a silicon nitride waveguide (blue dashed line).}
\end{figure}

\section{Cavity realisation}
Following Fig.~\ref{fig:sideband}(d), a cavity that sufficiently filters the sideband requires a larger $V_m$ than the design in Fig.~\ref{fig:phc}(b), to avoid strong ZPL splitting at $g \gg \kappa_c$. Motivated by this, we simulate a planar PhCC cell, shown in Fig.~\ref{fig:cav}(b), where, compared to Fig.~\ref{fig:phc}(b), there is no confinement in the z-direction, for which we find $V_m = 0.011 (\lambda/n)^3$. Removing the V-groove, the structure now has ease of fabrication, compatible with standard planar lithography, yet still provides a significantly smaller mode volume than standard designs.\cite{sipahigil2016integrated}. 

The cavity is realised by reducing the air hole diameter from 140 nm to 100 nm. By quadratically tapering the hole diameter, we form a Gaussian mode \cite{quan2010photonic}.  Fig.~\ref{fig:cav}(a) demonstrates the mode confinement of the cavity in the x-y plane, although in Fig.~\ref{fig:cav}(b) there is no longer strong confinement in the z-direction. To reach the desired $Q$, we use 10 mirror pairs and 5 taper pairs, each with a cell size of 250 nm, for a total cavity length below 10 $\mu$m. The total cavity mode volume is then $V_m = 0.075 (\lambda/n)^3$. As this cavity is shorter than high $Q$ designs it could be densely integrated at chip-scale. \cite{sipahigil2016integrated,hu2018experimental}.

To couple the NV centre it is necessary to model the effect of loading the silicon nitride cavity with a nanodiamond. We model this as a 20 nm cube with refractive index $n = 2.4$ at the centre of the cavity. In Fig.~\ref{fig:cav}(c)-(e), the nanodiamond acts to increase $V_m$ in the x-plane as it is larger than the 5 nm bow tie apex. However, in Fig.~\ref{fig:cav}(f)-(h), $V_m$ is reduced in the z-plane, owing to the introduced refractive index contrast. Through this trade-off, the overall $V_m$ is similar to the unloaded case. The low $Q$ of the structure is not greatly affected by the nanodiamond perturbation. The loaded cavity with five taper cells is near optimal for the 200 K parameter space with $ \beta I = 0.52$. 

In this moderate-$Q$ regime, the spectral response is calculated in FDTD using harmonic inversion \cite{mandelshtam1997harmonic}. The first two resonances of the TE excited structure are shown in Fig.~\ref{fig:cavspec}(a) with the first matching the ZPL at 470 THz.

To confirm the treatment of the cavity acting on the sideband as a filter, we collect the power in the waveguiding cross-section $\gamma_{wg}$ as a function of the dipole excitation frequency (see supplemental material for full details \cite{Note1}). Normalised by the dipole response in a homogeneous medium $\gamma_0$, we define the enhancement rate $ F_{P} = \gamma_{wg} / \gamma_0$ \cite{javadi2018numerical}. Fig \ref{fig:cavspec}(b) models a strong increase in emission at the first order resonance $ F_{P} > 50$. The higher order resonance at 775 THz lies above the light line in Fig.~\ref{fig:phc}(a), and hence is not guided. We observe a strong suppression of emission in the PhCC bandgap spanning the majority of the sideband with $ F_{P} \sim 0.1$, and a small increase in emission outside the bandgap corresponding to the tail of the sideband which we note is equivalent to the enhancement seen from a dipole in a silicon nitride waveguide $F_P \sim 2$. It is important to state this $F_P$ calculation neglects dephasing ($\gamma^*$ = 0), which would dominate the waveguide decay channel. Our approach to consider a sideband decay rate $ F_{P} =1$, compared to the strong enhancement of the ZPL, in a model that captures both dephasing and the sideband is thus justified as a worst case scenario. A further model could be developed to capture the dynamics of the sideband in the Master equation, considering the emission suppression in this calculation, which might result in even higher indistinguishabilities.

%In this moderate $Q$ regime, the response is calculated using harmonic inversion of the average response of point-like time domain monitors along the cavity cross-section.

\section{Tolerance in emitter--cavity coupling}
Given the stochastic formation of the atom-like emitter in the nanodiamond, variation in the orientation and location of the dipole with respect to the cavity should be considered as well as the effect of the fabrication induced variation on the cavity quality factor and mode volume. These considerations are important to realise an experimental demonstration of this system and to assess its viability as a reproducible and scalable source of indistinguishable photons.
 
\begin{figure}[b]
\includegraphics[width=\columnwidth]{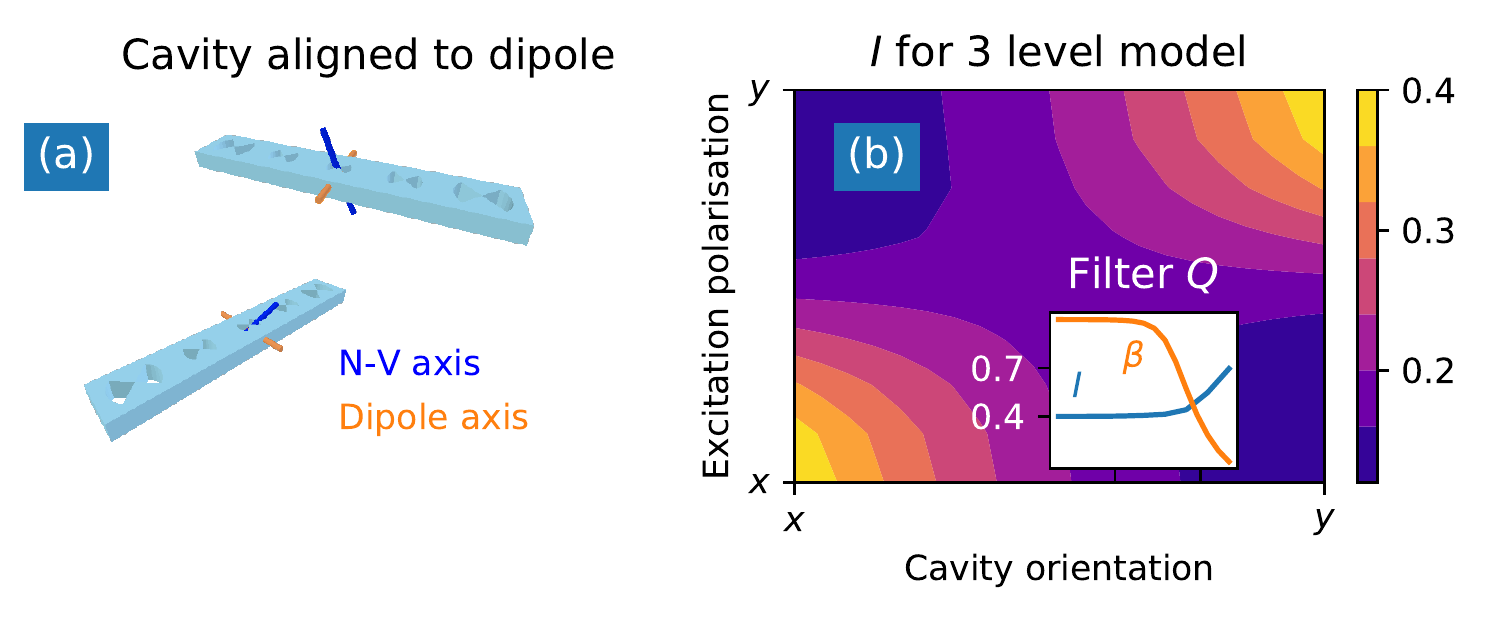}
\caption{\label{fig:cavpol}(a) Orientating the cavity with the projection onto the chip plane of the N-V axis (in blue) ensures good cavity coupling to the plane of optical dipoles (orange). (b) Indistinguishability at 200 K for a three level model including both dipoles. If polarising dephasing is dominant, coupling to a mixture of the two dipoles results in a decrease in indistinguishability, dependent on the excitation polarisation.}
\end{figure}

Most critically, the NV centre in a nanodiamond will be oriented at random with respect to the cavity polarisation. In this case, the rate $g$ is affected by the overlap of the dipole $a$ and the cavity field $c$ by the dot product $\epsilon^*_a\cdot\epsilon_c$ (see supplementary information for full expression \cite{Note1}). As the NV centre consists of two dipoles in a plane perpendicular to its N-V magnetic dipole, in the worst alignment, this plane is perpendicular to the cavity mode, resulting in an effective $g = 0$. Our strategy would be to find the orientation of the NV magnetic dipole and project this orientation into the plane of the sample. Shown in Fig.~\ref{fig:cavpol}(a), by orientating the cavity we can ensure $\epsilon^*_a\cdot\epsilon_c = 1$ for one of the two electric dipole axes of the emission. 

\begin{figure}
\includegraphics[width=\columnwidth]{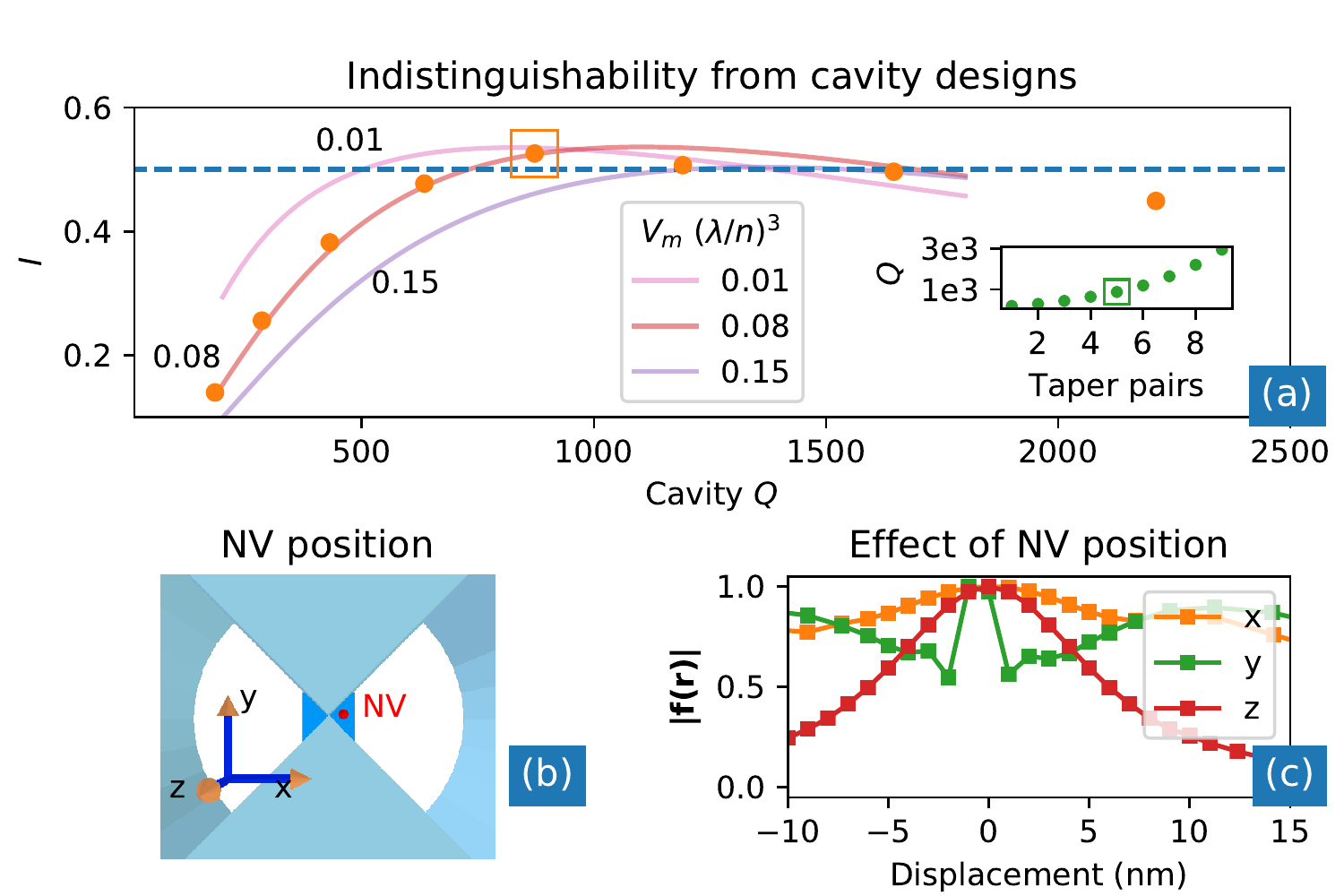}
\caption{\label{fig:cavtol}(a) The five taper cell design (highlighted by orange square and green square in inset) reaches optimal $I$ at the simulated $V_m$ and $Q$. (b) In realistic devices, the NV defect can be off-centre from the mode apex (c) Plotting the field structure $\mathbf{f}(r)$ models the effect of the NV centre position with respect to the cavity. }
\end{figure}

Another important consideration of the NV centre's level structure is that it consists of two orthogonally polarised dipoles, that have so far been treated independently by considering a single two level system with pure dephasing, as in previous cavity QED studies~\citep{albrecht2013coupling}. A fuller picture would consider transitions between these two excited states, and Abtew \emph{et al.}~\cite{abtew2011dynamic} identified the dynamic Jahn-Teller effect acting between these two states as the dominant dephasing mechanism in the NV centre. This is potentially important, as although the cavity outputs a single well-defined polarisation, the dynamics of the three level system could have significant effects on the spectral coherence. To explore the effect of this, we consider a three-level system with two excited states labelled $x$ and $y$. The master equation is modified to
\begin{align}
\partial_t &\rho(t) = -ig\qty[X(\theta),\rho(t)]+\gamma(\mathcal{L}_{\sigma_x}[\rho(t)]+\mathcal{L}_{\sigma_y}[\rho(t)])\nonumber\\
&+\gamma^*_{xy} \mathcal{L}_{\sigma_{xy}}[\rho(t)]+e^{-\Delta\beta}\gamma^*_{xy} \mathcal{L}_{\sigma_{yx}}[\rho(t)]+
\kappa_c \mathcal{L}_{a}[\rho(t)],
\end{align}
where $\sigma_x = |g\rangle\langle e_x|$ and $\sigma_y = |g\rangle\langle e_y|$ describe the decay from the corresponding excited state to the ground state, and $\sigma_{xy}$ and $\sigma_{yx}$ describe the respective transitions between the two excited states. We now have $X(\theta)=(\sigma_{m}(\theta)^{\dagger}a+\sigma_{m}(\theta)a^{\dagger})$ with $\sigma_m(\theta)=\sin(\theta)\sigma_x+\cos(\theta)\sigma_y$, and $\theta$ parameterises the orientation between the cavity mode and the dipole associated with $x$ excited state. Following Fu \emph{et al.}, strain induced splitting between the excited states is set at $\Delta$ = 100 GHz and at 200 K the polarisation relaxation rate between these two states is set at $\gamma^*_{xy} = 1$ THz \cite{fu2009observation}. In Fig \ref{fig:cavpol}(b) we plot the indistinguishability using this model as a function of the angle $\theta$ (cavity orientation) between the cavity and $x$ transition dipole, and the initial NV centre population set by the polarisation of the free space excitation $\rho_0(\theta) = ( \sin\theta|e_x\rangle\langle e_x| + \cos\theta|e_y\rangle\langle e_y| )  \otimes
|0\rangle\langle0|_{cav}$. Interestingly, the indistinguishability is reduced for all cavity orientations and initial states as compared to the two-level model, although can be recovered to $I$ > 0.7 through externally filtering. Moreover, this reduction is made worse when the initial state is not that to which the cavity couples. Since the cavity only couples to one dipole, the indistinguishability is reduced since now excitations can occupy states which are subject to dephasing for greater periods of time when compared to the two-level model, in which excitations always rapidly leave the NV centre. Further details of this three level model are included in the supplementary information \cite{Note1}.
%The second and third term describe spontaneous emission from each of the excited states $e_x$ and $e_y$ respectively. The fourth and fifth terms represent the decay/excitation between the two excited states. The last term describes spontaneous emission from the cavity as before. 

In relation to this potential reduction in indistinguishability, we note that Plakhotnik \emph{et al.} found an improved experimental fit to NV centre broadening at 200 K by considering pure dephasing $\gamma^* = 1$ THz via the intersystem crossing, which would dominate the polarisation dephasing at $\gamma^*_{xy} = 10$ GHz \cite{plakhotnik2015electron}, and give validity to our two level system model. It would be interesting in further work to combine pure dephasing mechanisms with polarisation dephasing, as the exact dephasing mechanism is still not well understood at this temperature regime~\cite{gali2019ab}. These considerations would then have to be combined with a cavity QED model, and the predictions could be experimentally explored by measuring the indistinguishability of photons emitted from different orientation NV centres in various cavity designs. Additionally, vector electric fields could be applied to reorient strain-induced linear dipoles with respect to the cavity orientation~\cite{acosta2012dynamic}.

Nanophotonics as fabricated have variations which can drastically affect function and this is typically problematic for high-$Q$ photonic cavities. Fig.~\ref{fig:cavtol}(a) compares different taper designs and indicates the 5 taper pair cavity is robust to 10 \% error in fabricated $Q$. In an experimental setting, it should be noted that the location of the NV centre within the nanodiamond affects the rate $g$ by the factor $\mathbf{f}(r)$ defining the spatial structure of the cavity electic field (see supplementary material for details \cite{Note1}). In Fig.~\ref{fig:cavtol}(c), we plot $\mathbf{f}(r)$ for the $V_m = 0.075$ cavity, noting that it falls to 0.5 at $\pm6$ nm from the mode maximum. Away from the maximum, we can consider an effective mode volume $V_m^* = V_m/|\mathbf{f}(r)|^2$. In Fig.~\ref{fig:cavtol}(a), $I < 0.5$ for the $V_m = 0.15$ contour, equivalent to $\mathbf{f}(r)$ = 0.7, indicating the NV centre position will be critical to realise good coupling.

\section{Conclusions}
In summary, we have proposed a class of ultra-small mode volume cavities that could significantly improve non-cryogenic emission of the NV centre for quantum information applications. By combining Peltier cooling with new PhCC structures, it is possible to increase the indistinguishability by nine orders of magnitude to $I = 0.54$. The Purcell enhancement results in photons effectively being emitted on demand under pulsed excitation. Here emission could be filtered to extract high purity indistinguishable photons with $I$ = 0.73 at \linebreak[4] 29 \% extraction efficiency. Higher values can be achieved with tighter filters or at lower temperatures if we relax the Peltier cooling limit. We could also potentially use this emission enhancement with single shot spin readout \cite{wolf2015purcell} and herald entanglement between remote spins. These cavity enhancement techniques could equally be applied to other systems where the phonon sideband is less pronounced~\cite{benedikter2017cavity,tran2017deterministic}, leading to highly indistinguishable emission at elevated temperatures, for linear optic quantum computation applications 
\cite{wang2017high}. In this work aimed at NV centres we have developed a planar nanodiamond-silicon nitride cavity design considering robustness to the $Q$ factor and the NV centre position. Following our recent work encapsulating NV centres in silicon nitride \cite{smith2019single}, it should be possible to fabricate this design as a step towards on-demand indistinguishable photons above cryogenic temperatures. 

\section*{Acknowledgements}
JS thanks Felipe Ortiz Huerta and Jorge Monroy Ruz for useful discussions. This work was supported by EPSRC (EP/L015544/1, EP/S023607/1, EP/M024458/1), ERC (SBS 3-5, 758843), and the British Council (IL 6 352345416).

\bibliography{bibl}% Produces the bibliography via BibTeX.

\begin{thebibliography}{30}%
\makeatletter
\providecommand \@ifxundefined [1]{%
 \@ifx{#1\undefined}
}%
\providecommand \@ifnum [1]{%
 \ifnum #1\expandafter \@firstoftwo
 \else \expandafter \@secondoftwo
 \fi
}%
\providecommand \@ifx [1]{%
 \ifx #1\expandafter \@firstoftwo
 \else \expandafter \@secondoftwo
 \fi
}%
\providecommand \natexlab [1]{#1}%
\providecommand \enquote  [1]{``#1''}%
\providecommand \bibnamefont  [1]{#1}%
\providecommand \bibfnamefont [1]{#1}%
\providecommand \citenamefont [1]{#1}%
\providecommand \href@noop [0]{\@secondoftwo}%
\providecommand \href [0]{\begingroup \@sanitize@url \@href}%
\providecommand \@href[1]{\@@startlink{#1}\@@href}%
\providecommand \@@href[1]{\endgroup#1\@@endlink}%
\providecommand \@sanitize@url [0]{\catcode `\\12\catcode `\$12\catcode
  `\&12\catcode `\#12\catcode `\^12\catcode `\_12\catcode `\%12\relax}%
\providecommand \@@startlink[1]{}%
\providecommand \@@endlink[0]{}%
\providecommand \url  [0]{\begingroup\@sanitize@url \@url }%
\providecommand \@url [1]{\endgroup\@href {#1}{\urlprefix }}%
\providecommand \urlprefix  [0]{URL }%
\providecommand \Eprint [0]{\href }%
\providecommand \doibase [0]{https://doi.org/}%
\providecommand \selectlanguage [0]{\@gobble}%
\providecommand \bibinfo  [0]{\@secondoftwo}%
\providecommand \bibfield  [0]{\@secondoftwo}%
\providecommand \translation [1]{[#1]}%
\providecommand \BibitemOpen [0]{}%
\providecommand \bibitemStop [0]{}%
\providecommand \bibitemNoStop [0]{.\EOS\space}%
\providecommand \EOS [0]{\spacefactor3000\relax}%
\providecommand \BibitemShut  [1]{\csname bibitem#1\endcsname}%
\let\auto@bib@innerbib\@empty
%</preamble>
\bibitem [{\citenamefont {Rozp{\k{e}}dek}\ \emph {et~al.}(2019)\citenamefont
  {Rozp{\k{e}}dek}, \citenamefont {Yehia}, \citenamefont {Goodenough},
  \citenamefont {Ruf}, \citenamefont {Humphreys}, \citenamefont {Hanson},
  \citenamefont {Wehner},\ and\ \citenamefont {Elkouss}}]{rozpkedek2019near}%
  \BibitemOpen
  \bibfield  {author} {\bibinfo {author} {\bibfnamefont {F.}~\bibnamefont
  {Rozp{\k{e}}dek}}, \bibinfo {author} {\bibfnamefont {R.}~\bibnamefont
  {Yehia}}, \bibinfo {author} {\bibfnamefont {K.}~\bibnamefont {Goodenough}},
  \bibinfo {author} {\bibfnamefont {M.}~\bibnamefont {Ruf}}, \bibinfo {author}
  {\bibfnamefont {P.~C.}\ \bibnamefont {Humphreys}}, \bibinfo {author}
  {\bibfnamefont {R.}~\bibnamefont {Hanson}}, \bibinfo {author} {\bibfnamefont
  {S.}~\bibnamefont {Wehner}},\ and\ \bibinfo {author} {\bibfnamefont
  {D.}~\bibnamefont {Elkouss}},\ }\bibfield  {title} {\bibinfo {title}
  {Near-term quantum-repeater experiments with nitrogen-vacancy centers:
  Overcoming the limitations of direct transmission},\ }\href@noop {}
  {\bibfield  {journal} {\bibinfo  {journal} {Physical Review A}\ }\textbf
  {\bibinfo {volume} {99}},\ \bibinfo {pages} {052330} (\bibinfo {year}
  {2019})}\BibitemShut {NoStop}%
\bibitem [{\citenamefont {Bernien}\ \emph {et~al.}(2013)\citenamefont
  {Bernien}, \citenamefont {Hensen}, \citenamefont {Pfaff}, \citenamefont
  {Koolstra}, \citenamefont {Blok}, \citenamefont {Robledo}, \citenamefont
  {Taminiau}, \citenamefont {Markham}, \citenamefont {Twitchen}, \citenamefont
  {Childress} \emph {et~al.}}]{bernien2013heralded}%
  \BibitemOpen
  \bibfield  {author} {\bibinfo {author} {\bibfnamefont {H.}~\bibnamefont
  {Bernien}}, \bibinfo {author} {\bibfnamefont {B.}~\bibnamefont {Hensen}},
  \bibinfo {author} {\bibfnamefont {W.}~\bibnamefont {Pfaff}}, \bibinfo
  {author} {\bibfnamefont {G.}~\bibnamefont {Koolstra}}, \bibinfo {author}
  {\bibfnamefont {M.}~\bibnamefont {Blok}}, \bibinfo {author} {\bibfnamefont
  {L.}~\bibnamefont {Robledo}}, \bibinfo {author} {\bibfnamefont
  {T.}~\bibnamefont {Taminiau}}, \bibinfo {author} {\bibfnamefont
  {M.}~\bibnamefont {Markham}}, \bibinfo {author} {\bibfnamefont
  {D.}~\bibnamefont {Twitchen}}, \bibinfo {author} {\bibfnamefont
  {L.}~\bibnamefont {Childress}}, \emph {et~al.},\ }\bibfield  {title}
  {\bibinfo {title} {Heralded entanglement between solid-state qubits separated
  by three metres},\ }\href@noop {} {\bibfield  {journal} {\bibinfo  {journal}
  {Nature}\ }\textbf {\bibinfo {volume} {497}},\ \bibinfo {pages} {86}
  (\bibinfo {year} {2013})}\BibitemShut {NoStop}%
\bibitem [{\citenamefont {Albrecht}\ \emph {et~al.}(2013)\citenamefont
  {Albrecht}, \citenamefont {Bommer}, \citenamefont {Deutsch}, \citenamefont
  {Reichel},\ and\ \citenamefont {Becher}}]{albrecht2013coupling}%
  \BibitemOpen
  \bibfield  {author} {\bibinfo {author} {\bibfnamefont {R.}~\bibnamefont
  {Albrecht}}, \bibinfo {author} {\bibfnamefont {A.}~\bibnamefont {Bommer}},
  \bibinfo {author} {\bibfnamefont {C.}~\bibnamefont {Deutsch}}, \bibinfo
  {author} {\bibfnamefont {J.}~\bibnamefont {Reichel}},\ and\ \bibinfo {author}
  {\bibfnamefont {C.}~\bibnamefont {Becher}},\ }\bibfield  {title} {\bibinfo
  {title} {Coupling of a single nitrogen-vacancy center in diamond to a
  fiber-based microcavity},\ }\href@noop {} {\bibfield  {journal} {\bibinfo
  {journal} {Physical review letters}\ }\textbf {\bibinfo {volume} {110}},\
  \bibinfo {pages} {243602} (\bibinfo {year} {2013})}\BibitemShut {NoStop}%
\bibitem [{\citenamefont {Iles-Smith}\ \emph {et~al.}(2017)\citenamefont
  {Iles-Smith}, \citenamefont {McCutcheon}, \citenamefont {Nazir},\ and\
  \citenamefont {M{\o}rk}}]{iles2017phonon}%
  \BibitemOpen
  \bibfield  {author} {\bibinfo {author} {\bibfnamefont {J.}~\bibnamefont
  {Iles-Smith}}, \bibinfo {author} {\bibfnamefont {D.~P.}\ \bibnamefont
  {McCutcheon}}, \bibinfo {author} {\bibfnamefont {A.}~\bibnamefont {Nazir}},\
  and\ \bibinfo {author} {\bibfnamefont {J.}~\bibnamefont {M{\o}rk}},\
  }\bibfield  {title} {\bibinfo {title} {Phonon scattering inhibits
  simultaneous near-unity efficiency and indistinguishability in semiconductor
  single-photon sources},\ }\href@noop {} {\bibfield  {journal} {\bibinfo
  {journal} {Nature Photonics}\ }\textbf {\bibinfo {volume} {11}},\ \bibinfo
  {pages} {521} (\bibinfo {year} {2017})}\BibitemShut {NoStop}%
\bibitem [{\citenamefont {Riedel}\ \emph {et~al.}(2017)\citenamefont {Riedel},
  \citenamefont {S{\"o}llner}, \citenamefont {Shields}, \citenamefont
  {Starosielec}, \citenamefont {Appel}, \citenamefont {Neu}, \citenamefont
  {Maletinsky},\ and\ \citenamefont {Warburton}}]{riedel2017deterministic}%
  \BibitemOpen
  \bibfield  {author} {\bibinfo {author} {\bibfnamefont {D.}~\bibnamefont
  {Riedel}}, \bibinfo {author} {\bibfnamefont {I.}~\bibnamefont {S{\"o}llner}},
  \bibinfo {author} {\bibfnamefont {B.~J.}\ \bibnamefont {Shields}}, \bibinfo
  {author} {\bibfnamefont {S.}~\bibnamefont {Starosielec}}, \bibinfo {author}
  {\bibfnamefont {P.}~\bibnamefont {Appel}}, \bibinfo {author} {\bibfnamefont
  {E.}~\bibnamefont {Neu}}, \bibinfo {author} {\bibfnamefont {P.}~\bibnamefont
  {Maletinsky}},\ and\ \bibinfo {author} {\bibfnamefont {R.~J.}\ \bibnamefont
  {Warburton}},\ }\bibfield  {title} {\bibinfo {title} {Deterministic
  enhancement of coherent photon generation from a nitrogen-vacancy center in
  ultrapure diamond},\ }\href@noop {} {\bibfield  {journal} {\bibinfo
  {journal} {Physical Review X}\ }\textbf {\bibinfo {volume} {7}},\ \bibinfo
  {pages} {031040} (\bibinfo {year} {2017})}\BibitemShut {NoStop}%
\bibitem [{\citenamefont {Chen}\ \emph {et~al.}(2017)\citenamefont {Chen},
  \citenamefont {Salter}, \citenamefont {Knauer}, \citenamefont {Weng},
  \citenamefont {Frangeskou}, \citenamefont {Stephen}, \citenamefont {Ishmael},
  \citenamefont {Dolan}, \citenamefont {Johnson}, \citenamefont {Green} \emph
  {et~al.}}]{chen2017laser}%
  \BibitemOpen
  \bibfield  {author} {\bibinfo {author} {\bibfnamefont {Y.-C.}\ \bibnamefont
  {Chen}}, \bibinfo {author} {\bibfnamefont {P.~S.}\ \bibnamefont {Salter}},
  \bibinfo {author} {\bibfnamefont {S.}~\bibnamefont {Knauer}}, \bibinfo
  {author} {\bibfnamefont {L.}~\bibnamefont {Weng}}, \bibinfo {author}
  {\bibfnamefont {A.~C.}\ \bibnamefont {Frangeskou}}, \bibinfo {author}
  {\bibfnamefont {C.~J.}\ \bibnamefont {Stephen}}, \bibinfo {author}
  {\bibfnamefont {S.~N.}\ \bibnamefont {Ishmael}}, \bibinfo {author}
  {\bibfnamefont {P.~R.}\ \bibnamefont {Dolan}}, \bibinfo {author}
  {\bibfnamefont {S.}~\bibnamefont {Johnson}}, \bibinfo {author} {\bibfnamefont
  {B.~L.}\ \bibnamefont {Green}}, \emph {et~al.},\ }\bibfield  {title}
  {\bibinfo {title} {Laser writing of coherent colour centres in diamond},\
  }\href@noop {} {\bibfield  {journal} {\bibinfo  {journal} {Nature Photonics}\
  }\textbf {\bibinfo {volume} {11}},\ \bibinfo {pages} {77} (\bibinfo {year}
  {2017})}\BibitemShut {NoStop}%
\bibitem [{\citenamefont {Kalb}\ \emph {et~al.}(2017)\citenamefont {Kalb},
  \citenamefont {Reiserer}, \citenamefont {Humphreys}, \citenamefont
  {Bakermans}, \citenamefont {Kamerling}, \citenamefont {Nickerson},
  \citenamefont {Benjamin}, \citenamefont {Twitchen}, \citenamefont {Markham},\
  and\ \citenamefont {Hanson}}]{kalb2017entanglement}%
  \BibitemOpen
  \bibfield  {author} {\bibinfo {author} {\bibfnamefont {N.}~\bibnamefont
  {Kalb}}, \bibinfo {author} {\bibfnamefont {A.~A.}\ \bibnamefont {Reiserer}},
  \bibinfo {author} {\bibfnamefont {P.~C.}\ \bibnamefont {Humphreys}}, \bibinfo
  {author} {\bibfnamefont {J.~J.}\ \bibnamefont {Bakermans}}, \bibinfo {author}
  {\bibfnamefont {S.~J.}\ \bibnamefont {Kamerling}}, \bibinfo {author}
  {\bibfnamefont {N.~H.}\ \bibnamefont {Nickerson}}, \bibinfo {author}
  {\bibfnamefont {S.~C.}\ \bibnamefont {Benjamin}}, \bibinfo {author}
  {\bibfnamefont {D.~J.}\ \bibnamefont {Twitchen}}, \bibinfo {author}
  {\bibfnamefont {M.}~\bibnamefont {Markham}},\ and\ \bibinfo {author}
  {\bibfnamefont {R.}~\bibnamefont {Hanson}},\ }\bibfield  {title} {\bibinfo
  {title} {Entanglement distillation between solid-state quantum network
  nodes},\ }\href@noop {} {\bibfield  {journal} {\bibinfo  {journal} {Science}\
  }\textbf {\bibinfo {volume} {356}},\ \bibinfo {pages} {928} (\bibinfo {year}
  {2017})}\BibitemShut {NoStop}%
\bibitem [{\citenamefont {Bogdanov}\ \emph {et~al.}(2019)\citenamefont
  {Bogdanov}, \citenamefont {Boltasseva},\ and\ \citenamefont
  {Shalaev}}]{bogdanov2019overcoming}%
  \BibitemOpen
  \bibfield  {author} {\bibinfo {author} {\bibfnamefont {S.~I.}\ \bibnamefont
  {Bogdanov}}, \bibinfo {author} {\bibfnamefont {A.}~\bibnamefont
  {Boltasseva}},\ and\ \bibinfo {author} {\bibfnamefont {V.~M.}\ \bibnamefont
  {Shalaev}},\ }\bibfield  {title} {\bibinfo {title} {Overcoming quantum
  decoherence with plasmonics},\ }\href@noop {} {\bibfield  {journal} {\bibinfo
   {journal} {Science}\ }\textbf {\bibinfo {volume} {364}},\ \bibinfo {pages}
  {532} (\bibinfo {year} {2019})}\BibitemShut {NoStop}%
\bibitem [{\citenamefont {Bogdanov}\ \emph {et~al.}(2020)\citenamefont
  {Bogdanov}, \citenamefont {Makarova}, \citenamefont {Xu}, \citenamefont
  {Martin}, \citenamefont {Lagutchev}, \citenamefont {Olinde}, \citenamefont
  {Shah}, \citenamefont {Chowdhury}, \citenamefont {Gabidullin}, \citenamefont
  {Ryzhikov} \emph {et~al.}}]{bogdanov2020ultrafast}%
  \BibitemOpen
  \bibfield  {author} {\bibinfo {author} {\bibfnamefont {S.~I.}\ \bibnamefont
  {Bogdanov}}, \bibinfo {author} {\bibfnamefont {O.~A.}\ \bibnamefont
  {Makarova}}, \bibinfo {author} {\bibfnamefont {X.}~\bibnamefont {Xu}},
  \bibinfo {author} {\bibfnamefont {Z.~O.}\ \bibnamefont {Martin}}, \bibinfo
  {author} {\bibfnamefont {A.~S.}\ \bibnamefont {Lagutchev}}, \bibinfo {author}
  {\bibfnamefont {M.}~\bibnamefont {Olinde}}, \bibinfo {author} {\bibfnamefont
  {D.}~\bibnamefont {Shah}}, \bibinfo {author} {\bibfnamefont {S.~N.}\
  \bibnamefont {Chowdhury}}, \bibinfo {author} {\bibfnamefont {A.~R.}\
  \bibnamefont {Gabidullin}}, \bibinfo {author} {\bibfnamefont {I.~A.}\
  \bibnamefont {Ryzhikov}}, \emph {et~al.},\ }\bibfield  {title} {\bibinfo
  {title} {Ultrafast quantum photonics enabled by coupling plasmonic
  nanocavities to strongly radiative antennas},\ }\href@noop {} {\bibfield
  {journal} {\bibinfo  {journal} {Optica}\ }\textbf {\bibinfo {volume} {7}},\
  \bibinfo {pages} {463} (\bibinfo {year} {2020})}\BibitemShut {NoStop}%
\bibitem [{\citenamefont {Wolf}\ \emph {et~al.}(2015)\citenamefont {Wolf},
  \citenamefont {Rosenberg}, \citenamefont {Rapaport},\ and\ \citenamefont
  {Bar-Gill}}]{wolf2015purcell}%
  \BibitemOpen
  \bibfield  {author} {\bibinfo {author} {\bibfnamefont {S.~A.}\ \bibnamefont
  {Wolf}}, \bibinfo {author} {\bibfnamefont {I.}~\bibnamefont {Rosenberg}},
  \bibinfo {author} {\bibfnamefont {R.}~\bibnamefont {Rapaport}},\ and\
  \bibinfo {author} {\bibfnamefont {N.}~\bibnamefont {Bar-Gill}},\ }\bibfield
  {title} {\bibinfo {title} {Purcell-enhanced optical spin readout of
  nitrogen-vacancy centers in diamond},\ }\href@noop {} {\bibfield  {journal}
  {\bibinfo  {journal} {Physical Review B}\ }\textbf {\bibinfo {volume} {92}},\
  \bibinfo {pages} {235410} (\bibinfo {year} {2015})}\BibitemShut {NoStop}%
\bibitem [{\citenamefont {Jung}\ \emph {et~al.}(2019)\citenamefont {Jung},
  \citenamefont {G{\"o}rlitz}, \citenamefont {Kambs}, \citenamefont {Pauly},
  \citenamefont {Raatz}, \citenamefont {Nelz}, \citenamefont {Neu},
  \citenamefont {Edmonds}, \citenamefont {Markham}, \citenamefont
  {M{\"u}cklich} \emph {et~al.}}]{jung2019spin}%
  \BibitemOpen
  \bibfield  {author} {\bibinfo {author} {\bibfnamefont {T.}~\bibnamefont
  {Jung}}, \bibinfo {author} {\bibfnamefont {J.}~\bibnamefont {G{\"o}rlitz}},
  \bibinfo {author} {\bibfnamefont {B.}~\bibnamefont {Kambs}}, \bibinfo
  {author} {\bibfnamefont {C.}~\bibnamefont {Pauly}}, \bibinfo {author}
  {\bibfnamefont {N.}~\bibnamefont {Raatz}}, \bibinfo {author} {\bibfnamefont
  {R.}~\bibnamefont {Nelz}}, \bibinfo {author} {\bibfnamefont {E.}~\bibnamefont
  {Neu}}, \bibinfo {author} {\bibfnamefont {A.~M.}\ \bibnamefont {Edmonds}},
  \bibinfo {author} {\bibfnamefont {M.}~\bibnamefont {Markham}}, \bibinfo
  {author} {\bibfnamefont {F.}~\bibnamefont {M{\"u}cklich}}, \emph {et~al.},\
  }\bibfield  {title} {\bibinfo {title} {Spin measurements of {NV} centers
  coupled to a photonic crystal cavity},\ }\href@noop {} {\bibfield  {journal}
  {\bibinfo  {journal} {arXiv preprint arXiv:1907.07602}\ } (\bibinfo {year}
  {2019})}\BibitemShut {NoStop}%
\bibitem [{\citenamefont {Hu}\ \emph {et~al.}(2018)\citenamefont {Hu},
  \citenamefont {Khater}, \citenamefont {Salas-Montiel}, \citenamefont
  {Kratschmer}, \citenamefont {Engelmann}, \citenamefont {Green},\ and\
  \citenamefont {Weiss}}]{hu2018experimental}%
  \BibitemOpen
  \bibfield  {author} {\bibinfo {author} {\bibfnamefont {S.}~\bibnamefont
  {Hu}}, \bibinfo {author} {\bibfnamefont {M.}~\bibnamefont {Khater}}, \bibinfo
  {author} {\bibfnamefont {R.}~\bibnamefont {Salas-Montiel}}, \bibinfo {author}
  {\bibfnamefont {E.}~\bibnamefont {Kratschmer}}, \bibinfo {author}
  {\bibfnamefont {S.}~\bibnamefont {Engelmann}}, \bibinfo {author}
  {\bibfnamefont {W.~M.}\ \bibnamefont {Green}},\ and\ \bibinfo {author}
  {\bibfnamefont {S.~M.}\ \bibnamefont {Weiss}},\ }\bibfield  {title} {\bibinfo
  {title} {Experimental realization of deep-subwavelength confinement in
  dielectric optical resonators},\ }\href@noop {} {\bibfield  {journal}
  {\bibinfo  {journal} {Science advances}\ }\textbf {\bibinfo {volume} {4}},\
  \bibinfo {pages} {eaat2355} (\bibinfo {year} {2018})}\BibitemShut {NoStop}%
\bibitem [{\citenamefont {Smith}\ \emph {et~al.}(2020)\citenamefont {Smith},
  \citenamefont {Monroy-Ruz}, \citenamefont {Rarity},\ and\ \citenamefont
  {C.~Balram}}]{smith2019single}%
  \BibitemOpen
  \bibfield  {author} {\bibinfo {author} {\bibfnamefont {J.}~\bibnamefont
  {Smith}}, \bibinfo {author} {\bibfnamefont {J.}~\bibnamefont {Monroy-Ruz}},
  \bibinfo {author} {\bibfnamefont {J.~G.}\ \bibnamefont {Rarity}},\ and\
  \bibinfo {author} {\bibfnamefont {K.}~\bibnamefont {C.~Balram}},\ }\bibfield
  {title} {\bibinfo {title} {Single photon emission and single spin coherence
  of a nitrogen vacancy center encapsulated in silicon nitride},\ }\href@noop
  {} {\bibfield  {journal} {\bibinfo  {journal} {Applied Physics Letters}\
  }\textbf {\bibinfo {volume} {116}},\ \bibinfo {pages} {134001} (\bibinfo
  {year} {2020})}\BibitemShut {NoStop}%
\bibitem [{\citenamefont {Kaer}\ \emph {et~al.}(2013)\citenamefont {Kaer},
  \citenamefont {Gregersen},\ and\ \citenamefont {Mork}}]{kaer2013role}%
  \BibitemOpen
  \bibfield  {author} {\bibinfo {author} {\bibfnamefont {P.}~\bibnamefont
  {Kaer}}, \bibinfo {author} {\bibfnamefont {N.}~\bibnamefont {Gregersen}},\
  and\ \bibinfo {author} {\bibfnamefont {J.}~\bibnamefont {Mork}},\ }\bibfield
  {title} {\bibinfo {title} {The role of phonon scattering in the
  indistinguishability of photons emitted from semiconductor cavity {QED}
  systems},\ }\href@noop {} {\bibfield  {journal} {\bibinfo  {journal} {New
  Journal of Physics}\ }\textbf {\bibinfo {volume} {15}},\ \bibinfo {pages}
  {035027} (\bibinfo {year} {2013})}\BibitemShut {NoStop}%
\bibitem [{\citenamefont {Wein}\ \emph {et~al.}(2018)\citenamefont {Wein},
  \citenamefont {Lauk}, \citenamefont {Ghobadi},\ and\ \citenamefont
  {Simon}}]{wein2018feasibility}%
  \BibitemOpen
  \bibfield  {author} {\bibinfo {author} {\bibfnamefont {S.}~\bibnamefont
  {Wein}}, \bibinfo {author} {\bibfnamefont {N.}~\bibnamefont {Lauk}}, \bibinfo
  {author} {\bibfnamefont {R.}~\bibnamefont {Ghobadi}},\ and\ \bibinfo {author}
  {\bibfnamefont {C.}~\bibnamefont {Simon}},\ }\bibfield  {title} {\bibinfo
  {title} {Feasibility of efficient room-temperature solid-state sources of
  indistinguishable single photons using ultrasmall mode volume cavities},\
  }\href@noop {} {\bibfield  {journal} {\bibinfo  {journal} {Physical Review
  B}\ }\textbf {\bibinfo {volume} {97}},\ \bibinfo {pages} {205418} (\bibinfo
  {year} {2018})}\BibitemShut {NoStop}%
\bibitem [{\citenamefont {Sipahigil}\ \emph {et~al.}(2016)\citenamefont
  {Sipahigil}, \citenamefont {Evans}, \citenamefont {Sukachev}, \citenamefont
  {Burek}, \citenamefont {Borregaard}, \citenamefont {Bhaskar}, \citenamefont
  {Nguyen}, \citenamefont {Pacheco}, \citenamefont {Atikian}, \citenamefont
  {Meuwly} \emph {et~al.}}]{sipahigil2016integrated}%
  \BibitemOpen
  \bibfield  {author} {\bibinfo {author} {\bibfnamefont {A.}~\bibnamefont
  {Sipahigil}}, \bibinfo {author} {\bibfnamefont {R.~E.}\ \bibnamefont
  {Evans}}, \bibinfo {author} {\bibfnamefont {D.~D.}\ \bibnamefont {Sukachev}},
  \bibinfo {author} {\bibfnamefont {M.~J.}\ \bibnamefont {Burek}}, \bibinfo
  {author} {\bibfnamefont {J.}~\bibnamefont {Borregaard}}, \bibinfo {author}
  {\bibfnamefont {M.~K.}\ \bibnamefont {Bhaskar}}, \bibinfo {author}
  {\bibfnamefont {C.~T.}\ \bibnamefont {Nguyen}}, \bibinfo {author}
  {\bibfnamefont {J.~L.}\ \bibnamefont {Pacheco}}, \bibinfo {author}
  {\bibfnamefont {H.~A.}\ \bibnamefont {Atikian}}, \bibinfo {author}
  {\bibfnamefont {C.}~\bibnamefont {Meuwly}}, \emph {et~al.},\ }\bibfield
  {title} {\bibinfo {title} {An integrated diamond nanophotonics platform for
  quantum-optical networks},\ }\href@noop {} {\bibfield  {journal} {\bibinfo
  {journal} {Science}\ }\textbf {\bibinfo {volume} {354}},\ \bibinfo {pages}
  {847} (\bibinfo {year} {2016})}\BibitemShut {NoStop}%
\bibitem [{Note1()}]{Note1}%
  \BibitemOpen
  \bibinfo {note} {See Supplemental Material for theoretical and simulation
  detail to support the main text.}\BibitemShut {Stop}%
\bibitem [{\citenamefont {Grange}\ \emph {et~al.}(2015)\citenamefont {Grange},
  \citenamefont {Hornecker}, \citenamefont {Hunger}, \citenamefont {Poizat},
  \citenamefont {G{\'e}rard}, \citenamefont {Senellart},\ and\ \citenamefont
  {Auff{\`e}ves}}]{grange2015cavity}%
  \BibitemOpen
  \bibfield  {author} {\bibinfo {author} {\bibfnamefont {T.}~\bibnamefont
  {Grange}}, \bibinfo {author} {\bibfnamefont {G.}~\bibnamefont {Hornecker}},
  \bibinfo {author} {\bibfnamefont {D.}~\bibnamefont {Hunger}}, \bibinfo
  {author} {\bibfnamefont {J.-P.}\ \bibnamefont {Poizat}}, \bibinfo {author}
  {\bibfnamefont {J.-M.}\ \bibnamefont {G{\'e}rard}}, \bibinfo {author}
  {\bibfnamefont {P.}~\bibnamefont {Senellart}},\ and\ \bibinfo {author}
  {\bibfnamefont {A.}~\bibnamefont {Auff{\`e}ves}},\ }\bibfield  {title}
  {\bibinfo {title} {Cavity-funneled generation of indistinguishable single
  photons from strongly dissipative quantum emitters},\ }\href@noop {}
  {\bibfield  {journal} {\bibinfo  {journal} {Physical review letters}\
  }\textbf {\bibinfo {volume} {114}},\ \bibinfo {pages} {193601} (\bibinfo
  {year} {2015})}\BibitemShut {NoStop}%
\bibitem [{\citenamefont {Fu}\ \emph {et~al.}(2009)\citenamefont {Fu},
  \citenamefont {Santori}, \citenamefont {Barclay}, \citenamefont {Rogers},
  \citenamefont {Manson},\ and\ \citenamefont
  {Beausoleil}}]{fu2009observation}%
  \BibitemOpen
  \bibfield  {author} {\bibinfo {author} {\bibfnamefont {K.-M.~C.}\
  \bibnamefont {Fu}}, \bibinfo {author} {\bibfnamefont {C.}~\bibnamefont
  {Santori}}, \bibinfo {author} {\bibfnamefont {P.~E.}\ \bibnamefont
  {Barclay}}, \bibinfo {author} {\bibfnamefont {L.~J.}\ \bibnamefont {Rogers}},
  \bibinfo {author} {\bibfnamefont {N.~B.}\ \bibnamefont {Manson}},\ and\
  \bibinfo {author} {\bibfnamefont {R.~G.}\ \bibnamefont {Beausoleil}},\
  }\bibfield  {title} {\bibinfo {title} {Observation of the dynamic
  {J}ahn-{T}eller effect in the excited states of nitrogen-vacancy centers in
  diamond},\ }\href@noop {} {\bibfield  {journal} {\bibinfo  {journal}
  {Physical Review Letters}\ }\textbf {\bibinfo {volume} {103}},\ \bibinfo
  {pages} {256404} (\bibinfo {year} {2009})}\BibitemShut {NoStop}%
\bibitem [{\citenamefont {Deng}\ \emph {et~al.}(2019)\citenamefont {Deng},
  \citenamefont {Wang}, \citenamefont {Ding}, \citenamefont {Duan},
  \citenamefont {Qin}, \citenamefont {Chen}, \citenamefont {He}, \citenamefont
  {He}, \citenamefont {Li}, \citenamefont {Li} \emph
  {et~al.}}]{deng2019quantum}%
  \BibitemOpen
  \bibfield  {author} {\bibinfo {author} {\bibfnamefont {Y.-H.}\ \bibnamefont
  {Deng}}, \bibinfo {author} {\bibfnamefont {H.}~\bibnamefont {Wang}}, \bibinfo
  {author} {\bibfnamefont {X.}~\bibnamefont {Ding}}, \bibinfo {author}
  {\bibfnamefont {Z.-C.}\ \bibnamefont {Duan}}, \bibinfo {author}
  {\bibfnamefont {J.}~\bibnamefont {Qin}}, \bibinfo {author} {\bibfnamefont
  {M.-C.}\ \bibnamefont {Chen}}, \bibinfo {author} {\bibfnamefont
  {Y.}~\bibnamefont {He}}, \bibinfo {author} {\bibfnamefont {Y.-M.}\
  \bibnamefont {He}}, \bibinfo {author} {\bibfnamefont {J.-P.}\ \bibnamefont
  {Li}}, \bibinfo {author} {\bibfnamefont {Y.-H.}\ \bibnamefont {Li}}, \emph
  {et~al.},\ }\bibfield  {title} {\bibinfo {title} {Quantum interference
  between light sources separated by 150 million kilometers},\ }\href@noop {}
  {\bibfield  {journal} {\bibinfo  {journal} {Physical review letters}\
  }\textbf {\bibinfo {volume} {123}},\ \bibinfo {pages} {080401} (\bibinfo
  {year} {2019})}\BibitemShut {NoStop}%
\bibitem [{\citenamefont {Quan}\ \emph {et~al.}(2010)\citenamefont {Quan},
  \citenamefont {Deotare},\ and\ \citenamefont {Loncar}}]{quan2010photonic}%
  \BibitemOpen
  \bibfield  {author} {\bibinfo {author} {\bibfnamefont {Q.}~\bibnamefont
  {Quan}}, \bibinfo {author} {\bibfnamefont {P.~B.}\ \bibnamefont {Deotare}},\
  and\ \bibinfo {author} {\bibfnamefont {M.}~\bibnamefont {Loncar}},\
  }\bibfield  {title} {\bibinfo {title} {Photonic crystal nanobeam cavity
  strongly coupled to the feeding waveguide},\ }\href@noop {} {\bibfield
  {journal} {\bibinfo  {journal} {Applied Physics Letters}\ }\textbf {\bibinfo
  {volume} {96}},\ \bibinfo {pages} {203102} (\bibinfo {year}
  {2010})}\BibitemShut {NoStop}%
\bibitem [{\citenamefont {Mandelshtam}\ and\ \citenamefont
  {Taylor}(1997)}]{mandelshtam1997harmonic}%
  \BibitemOpen
  \bibfield  {author} {\bibinfo {author} {\bibfnamefont {V.~A.}\ \bibnamefont
  {Mandelshtam}}\ and\ \bibinfo {author} {\bibfnamefont {H.~S.}\ \bibnamefont
  {Taylor}},\ }\bibfield  {title} {\bibinfo {title} {Harmonic inversion of time
  signals and its applications},\ }\href@noop {} {\bibfield  {journal}
  {\bibinfo  {journal} {The Journal of chemical physics}\ }\textbf {\bibinfo
  {volume} {107}},\ \bibinfo {pages} {6756} (\bibinfo {year}
  {1997})}\BibitemShut {NoStop}%
\bibitem [{\citenamefont {Javadi}\ \emph {et~al.}(2018)\citenamefont {Javadi},
  \citenamefont {Mahmoodian}, \citenamefont {S{\"o}llner},\ and\ \citenamefont
  {Lodahl}}]{javadi2018numerical}%
  \BibitemOpen
  \bibfield  {author} {\bibinfo {author} {\bibfnamefont {A.}~\bibnamefont
  {Javadi}}, \bibinfo {author} {\bibfnamefont {S.}~\bibnamefont {Mahmoodian}},
  \bibinfo {author} {\bibfnamefont {I.}~\bibnamefont {S{\"o}llner}},\ and\
  \bibinfo {author} {\bibfnamefont {P.}~\bibnamefont {Lodahl}},\ }\bibfield
  {title} {\bibinfo {title} {Numerical modeling of the coupling efficiency of
  single quantum emitters in photonic-crystal waveguides},\ }\href@noop {}
  {\bibfield  {journal} {\bibinfo  {journal} {JOSA B}\ }\textbf {\bibinfo
  {volume} {35}},\ \bibinfo {pages} {514} (\bibinfo {year} {2018})}\BibitemShut
  {NoStop}%
\bibitem [{\citenamefont {Abtew}\ \emph {et~al.}(2011)\citenamefont {Abtew},
  \citenamefont {Sun}, \citenamefont {Shih}, \citenamefont {Dev}, \citenamefont
  {Zhang},\ and\ \citenamefont {Zhang}}]{abtew2011dynamic}%
  \BibitemOpen
  \bibfield  {author} {\bibinfo {author} {\bibfnamefont {T.~A.}\ \bibnamefont
  {Abtew}}, \bibinfo {author} {\bibfnamefont {Y.}~\bibnamefont {Sun}}, \bibinfo
  {author} {\bibfnamefont {B.-C.}\ \bibnamefont {Shih}}, \bibinfo {author}
  {\bibfnamefont {P.}~\bibnamefont {Dev}}, \bibinfo {author} {\bibfnamefont
  {S.}~\bibnamefont {Zhang}},\ and\ \bibinfo {author} {\bibfnamefont
  {P.}~\bibnamefont {Zhang}},\ }\bibfield  {title} {\bibinfo {title} {Dynamic
  jahn-teller effect in the nv- center in diamond},\ }\href@noop {} {\bibfield
  {journal} {\bibinfo  {journal} {Physical review letters}\ }\textbf {\bibinfo
  {volume} {107}},\ \bibinfo {pages} {146403} (\bibinfo {year}
  {2011})}\BibitemShut {NoStop}%
\bibitem [{\citenamefont {Plakhotnik}\ \emph {et~al.}(2015)\citenamefont
  {Plakhotnik}, \citenamefont {Doherty},\ and\ \citenamefont
  {Manson}}]{plakhotnik2015electron}%
  \BibitemOpen
  \bibfield  {author} {\bibinfo {author} {\bibfnamefont {T.}~\bibnamefont
  {Plakhotnik}}, \bibinfo {author} {\bibfnamefont {M.~W.}\ \bibnamefont
  {Doherty}},\ and\ \bibinfo {author} {\bibfnamefont {N.~B.}\ \bibnamefont
  {Manson}},\ }\bibfield  {title} {\bibinfo {title} {Electron-phonon processes
  of the nitrogen-vacancy center in diamond},\ }\href@noop {} {\bibfield
  {journal} {\bibinfo  {journal} {Physical Review B}\ }\textbf {\bibinfo
  {volume} {92}},\ \bibinfo {pages} {081203} (\bibinfo {year}
  {2015})}\BibitemShut {NoStop}%
\bibitem [{\citenamefont {Gali}(2019)}]{gali2019ab}%
  \BibitemOpen
  \bibfield  {author} {\bibinfo {author} {\bibfnamefont {{\'A}.}~\bibnamefont
  {Gali}},\ }\bibfield  {title} {\bibinfo {title} {Ab initio theory of the
  nitrogen-vacancy center in diamond},\ }\href@noop {} {\bibfield  {journal}
  {\bibinfo  {journal} {Nanophotonics}\ }\textbf {\bibinfo {volume} {8}},\
  \bibinfo {pages} {1907} (\bibinfo {year} {2019})}\BibitemShut {NoStop}%
\bibitem [{\citenamefont {Acosta}\ \emph {et~al.}(2012)\citenamefont {Acosta},
  \citenamefont {Santori}, \citenamefont {Faraon}, \citenamefont {Huang},
  \citenamefont {Fu}, \citenamefont {Stacey}, \citenamefont {Simpson},
  \citenamefont {Ganesan}, \citenamefont {Tomljenovic-Hanic}, \citenamefont
  {Greentree} \emph {et~al.}}]{acosta2012dynamic}%
  \BibitemOpen
  \bibfield  {author} {\bibinfo {author} {\bibfnamefont {V.}~\bibnamefont
  {Acosta}}, \bibinfo {author} {\bibfnamefont {C.}~\bibnamefont {Santori}},
  \bibinfo {author} {\bibfnamefont {A.}~\bibnamefont {Faraon}}, \bibinfo
  {author} {\bibfnamefont {Z.}~\bibnamefont {Huang}}, \bibinfo {author}
  {\bibfnamefont {K.-M.}\ \bibnamefont {Fu}}, \bibinfo {author} {\bibfnamefont
  {A.}~\bibnamefont {Stacey}}, \bibinfo {author} {\bibfnamefont
  {D.}~\bibnamefont {Simpson}}, \bibinfo {author} {\bibfnamefont
  {K.}~\bibnamefont {Ganesan}}, \bibinfo {author} {\bibfnamefont
  {S.}~\bibnamefont {Tomljenovic-Hanic}}, \bibinfo {author} {\bibfnamefont
  {A.}~\bibnamefont {Greentree}}, \emph {et~al.},\ }\bibfield  {title}
  {\bibinfo {title} {Dynamic stabilization of the optical resonances of single
  nitrogen-vacancy centers in diamond},\ }\href@noop {} {\bibfield  {journal}
  {\bibinfo  {journal} {Physical review letters}\ }\textbf {\bibinfo {volume}
  {108}},\ \bibinfo {pages} {206401} (\bibinfo {year} {2012})}\BibitemShut
  {NoStop}%
\bibitem [{\citenamefont {Benedikter}\ \emph {et~al.}(2017)\citenamefont
  {Benedikter}, \citenamefont {Kaupp}, \citenamefont {H{\"u}mmer},
  \citenamefont {Liang}, \citenamefont {Bommer}, \citenamefont {Becher},
  \citenamefont {Krueger}, \citenamefont {Smith}, \citenamefont {H{\"a}nsch},\
  and\ \citenamefont {Hunger}}]{benedikter2017cavity}%
  \BibitemOpen
  \bibfield  {author} {\bibinfo {author} {\bibfnamefont {J.}~\bibnamefont
  {Benedikter}}, \bibinfo {author} {\bibfnamefont {H.}~\bibnamefont {Kaupp}},
  \bibinfo {author} {\bibfnamefont {T.}~\bibnamefont {H{\"u}mmer}}, \bibinfo
  {author} {\bibfnamefont {Y.}~\bibnamefont {Liang}}, \bibinfo {author}
  {\bibfnamefont {A.}~\bibnamefont {Bommer}}, \bibinfo {author} {\bibfnamefont
  {C.}~\bibnamefont {Becher}}, \bibinfo {author} {\bibfnamefont
  {A.}~\bibnamefont {Krueger}}, \bibinfo {author} {\bibfnamefont {J.~M.}\
  \bibnamefont {Smith}}, \bibinfo {author} {\bibfnamefont {T.~W.}\ \bibnamefont
  {H{\"a}nsch}},\ and\ \bibinfo {author} {\bibfnamefont {D.}~\bibnamefont
  {Hunger}},\ }\bibfield  {title} {\bibinfo {title} {Cavity-enhanced
  single-photon source based on the silicon-vacancy center in diamond},\
  }\href@noop {} {\bibfield  {journal} {\bibinfo  {journal} {Physical Review
  Applied}\ }\textbf {\bibinfo {volume} {7}},\ \bibinfo {pages} {024031}
  (\bibinfo {year} {2017})}\BibitemShut {NoStop}%
\bibitem [{\citenamefont {Tran}\ \emph {et~al.}(2017)\citenamefont {Tran},
  \citenamefont {Wang}, \citenamefont {Xu}, \citenamefont {Yang}, \citenamefont
  {Toth}, \citenamefont {Odom},\ and\ \citenamefont
  {Aharonovich}}]{tran2017deterministic}%
  \BibitemOpen
  \bibfield  {author} {\bibinfo {author} {\bibfnamefont {T.~T.}\ \bibnamefont
  {Tran}}, \bibinfo {author} {\bibfnamefont {D.}~\bibnamefont {Wang}}, \bibinfo
  {author} {\bibfnamefont {Z.-Q.}\ \bibnamefont {Xu}}, \bibinfo {author}
  {\bibfnamefont {A.}~\bibnamefont {Yang}}, \bibinfo {author} {\bibfnamefont
  {M.}~\bibnamefont {Toth}}, \bibinfo {author} {\bibfnamefont {T.~W.}\
  \bibnamefont {Odom}},\ and\ \bibinfo {author} {\bibfnamefont
  {I.}~\bibnamefont {Aharonovich}},\ }\bibfield  {title} {\bibinfo {title}
  {Deterministic coupling of quantum emitters in 2{D} materials to plasmonic
  nanocavity arrays},\ }\href@noop {} {\bibfield  {journal} {\bibinfo
  {journal} {Nano letters}\ }\textbf {\bibinfo {volume} {17}},\ \bibinfo
  {pages} {2634} (\bibinfo {year} {2017})}\BibitemShut {NoStop}%
\bibitem [{\citenamefont {Wang}\ \emph {et~al.}(2017)\citenamefont {Wang},
  \citenamefont {He}, \citenamefont {Li}, \citenamefont {Su}, \citenamefont
  {Li}, \citenamefont {Huang}, \citenamefont {Ding}, \citenamefont {Chen},
  \citenamefont {Liu}, \citenamefont {Qin} \emph {et~al.}}]{wang2017high}%
  \BibitemOpen
  \bibfield  {author} {\bibinfo {author} {\bibfnamefont {H.}~\bibnamefont
  {Wang}}, \bibinfo {author} {\bibfnamefont {Y.}~\bibnamefont {He}}, \bibinfo
  {author} {\bibfnamefont {Y.-H.}\ \bibnamefont {Li}}, \bibinfo {author}
  {\bibfnamefont {Z.-E.}\ \bibnamefont {Su}}, \bibinfo {author} {\bibfnamefont
  {B.}~\bibnamefont {Li}}, \bibinfo {author} {\bibfnamefont {H.-L.}\
  \bibnamefont {Huang}}, \bibinfo {author} {\bibfnamefont {X.}~\bibnamefont
  {Ding}}, \bibinfo {author} {\bibfnamefont {M.-C.}\ \bibnamefont {Chen}},
  \bibinfo {author} {\bibfnamefont {C.}~\bibnamefont {Liu}}, \bibinfo {author}
  {\bibfnamefont {J.}~\bibnamefont {Qin}}, \emph {et~al.},\ }\bibfield  {title}
  {\bibinfo {title} {High-efficiency multiphoton boson sampling},\ }\href@noop
  {} {\bibfield  {journal} {\bibinfo  {journal} {Nature Photonics}\ }\textbf
  {\bibinfo {volume} {11}},\ \bibinfo {pages} {361} (\bibinfo {year}
  {2017})}\BibitemShut {NoStop}%
\end{thebibliography}%



\end{document}